# A Novel Approach to Discover Switch Behaviours in Process Mining

Yang Lu[0000-0002-9002-8650], Qifan Chen[0000-0003-1068-6408] and Simon Poon[0000-0003-2726-9109]

School of Computer Science, The University of Sydney, Sydney, NSW 2006, Australia
{yalu8986, qche8411}@uni.sydney.edu.au, simon.poon@sydney.edu.au

**Abstract.** Process mining is a relatively new subject which builds a bridge between process modelling and data mining. An exclusive choice in a process model usually splits the process into different branches. However, in some processes, it is possible to switch from one branch to another. The inductive miner guarantees to return sound process models, but fails to return a precise model when there are switch behaviours between different exclusive choice branches due to the limitation of process trees. In this paper, we present a novel extension to the process tree model to support switch behaviours between different branches of the exclusive choice operator and propose a novel extension to the inductive miner to discover sound process models with switch behaviours. The proposed discovery technique utilizes the theory of a previous study to detect possible switch behaviours. We apply both artificial and publicly-available datasets to evaluate our approach. Our results show that our approach can improve the precision of discovered models by 36% while maintaining high fitness values compared to the original inductive miner.

**Keywords:** Process Discovery, Complex Behaviours Detection, Switch Behaviours, Inductive Miner, Process Trees.

## 1 Introduction

Process mining is useful for analyzing business processes along with improving and predicting which contains three parts – process discovery, conformance checking and process enhancement [1]. The most critical part of process mining is process discovery, which aims at extracting insight of the system workflow from real data. The resulting process model should not only have a high fitness value, but also be an accurate representation of the real process [2]. The inductive miner is one of the leading process discovery algorithms which can guarantee to produce sound process models within finite time [1, 3]. Given the direct outcome of the inductive miner is a process tree [3], the behaviours being represented are limited. When giving complex event logs as input, the inductive miner often returns so-called "flower models" which preserve high fitness but have very low precision values [3, 4]. Although we can still replay the majority of traces on the process model, "flower models" fail to represent real processes accurately and precisely [1, 2].

When dealing with an exclusive decision choice in a process model, the decision point is split into multiple branches [5]. However, in many real-life processes, it can be possible to switch between branches after a decision has been made. Although the inductive miner is known to be useful in generating sound models from data, it fails to discover an accurate model when switch behaviours exist.

In this paper, we propose a novel extension to the process tree model to handle switch behaviours between different exclusive choice branches. We then develop a novel extension to the inductive miner to discover sound process models with switch behaviours based on the theory in [6]. From a broader perspective, our proposed method not only guarantees to produce sound process models but also not being limited to produce block-structured process models. We apply both artificial and publicly-available datasets to evaluate our approach. Fitness, precision and F-score [4, 7] are used to measure the accuracy of resulting models, size (the number of nodes) and CFC (the number of branching caused by split gateways) [8] are adopted to measure the model complexity.

The rest of the paper is structured as follows: Section 2 is a literature review of related work. Section 3 introduces formal definitions of some terms. Section 4 introduces the extension to the process tree model and how to translate it into a workflow net. In Section 5, we describe our process discovery technique. The approach is evaluated in Section 6. We finally conclude our paper in Section 7.

## 2　Background

When modelling switch behaviours between different exclusive choice branches using Petri-nets, a hidden transition is needed since we cannot connect two places directly [6]. The classical alpha algorithm [9] cannot discover any hidden transitions. [6, 10] improve the classical alpha algorithm to allow the detection of invisible tasks. Although the alpha algorithms are not robust to noises and cannot guarantee to produce sound models. [6] proposes a heuristic for detecting invisible transitions between activities directly from event logs. If there is a hidden transition between two activities on different exclusive choice branches, a switch behaviour is detected.

In reviewing other process discovery algorithms which can discover switch behaviours between different exclusive choice branches including the alpha algorithms with invisible tasks [6, 10], heuristics miners [11], genetic miners [12] and the ILP algorithm [13], none of them can guarantee to produce a sound process model. In addition, some of them cannot handle noises, thus, not suitable to be applied to real data. Although the split miner [7] can discover switch behaviours and guarantee to produce deadlock-free models. It still cannot guarantee to produce sound models as defined in [9], which defines soundness as (a) safeness, (b) proper completion, (c) option to complete, (d) absence of dead tasks.

The inductive miners are a family of process discovery algorithms which utilize the divide-and-conquer approach in the field of process discovery [3, 14-18]. The inductive miners recursively divide the activities into different partitions and split event logs until base cases are touched. The direct outcomes of the inductive miners are process trees, which can be easily translated into equivalent block-structured workflow nets [3]. An important feature of the inductive miner family is that the resulting model is always sound regardless of the input log. However, process trees also limit the behaviours which can be represented. For example, they fail to represent switch behaviours between exclusive choice branches.

When the given event log is complex, the inductive miner [3] can easily return a "flower model" with high fitness but low precision. [14] removes infrequent relations between activities before partitioning the activities. However, according to the benchmark in both [4] and [7]. The inductive miner infrequent (IMf) in [14] still returns models with low precision values compared with other algorithms. [19] tries to solve the problem by giving duplicate labels to the same activity when a local "flower model" is returned. The algorithm successfully improves the precision of the outcome models but leads to longer execution time. Besides, if we apply the algorithm in [19] with the inductive miners, the outcome models are still block-structured workflow nets.

The process mining framework in the original inductive miner [3] allows researchers to define their ways to partition activities and customized process tree semantics. For example, [17] puts lifecycle information into the process discovery to distinguish "interleaving" behaviours from "parallel" behaviours. [18] defines new operators on the process tree and uses the inductive miner to discover cancellation behaviours.

## 3      Preliminaries

In this section, we present some formal definitions which will be used in this paper. For process trees and block-structured workflow nets, we refer to [3], for soundness of Petri-nets, we refer to [9]. Besides, for IWF-net (workflow nets with invisible tasks), DIWF-net (a subset of IWF-nets) and log completeness, we refer to [6]. For clarification, in this paper, we use "X" to represent the exclusive choice operator, "→" to represent the sequence operator, "∧" to represent the parallel operator and "↺" to represent the loop operator in the process tree [3].

**Definition 1 (Relations between activities).** Let L be an event log of a workflow net N, let $a, b$ be two activities in L. Then:
1. $a >_L b$ if there is a trace $t \in L$ where t = <……, $a, b,$ …… >,
2. $a \sim_L b$ if there is a trace $t \in L$ where t = <……, $a, b, a,$ …… >, and there is a trace $t \in L$ where t = <……, $b, a, b,$ …… >,
3. $a \rightarrow_L b$ if $a >_L b \wedge (b \not>_L a \vee a \sim_L b)$,
4. $a \|_L b$ if $a >_L b \wedge b >_L a \wedge a \not\sim_L b$.

**Definition 2 (Mendacious dependency) [6].** Let N = (P, $T_v \cup T_{iv}$, F) be a potential sound IWF-net, $T_v$ is the set of visible tasks, $T_{iv}$ is the set of invisible tasks. There is a mendacious dependency between activities $a, b$ in event log L, denoted as $a \rightsquigarrow_L b$, iff $a \rightarrow_L b \wedge \exists x, y \in T_v: a \rightarrow_L x \wedge y \rightarrow_L b \wedge y \not>_L x \wedge x \not\|_L b \wedge a \not\|_L y$.

## 4 The Switch Process Tree

In this section, we formally define the switch behaviour and its corresponding representation on the process tree. The switch process tree is a novel extension based on the process tree model described in [3].

**Definition 3 (First, Path function).** Let $n$ be a leaf in a process tree, $tp$ be an arbitrary operator type. First $(n, tp)$ refers to the first ancestor node of $n$ with operator type $tp$. For example, in the process tree shown in Fig. 1, First (Node 3, X) refers to the root node. First $(n, tp) = \emptyset$ if none of the ancestor nodes of $n$ has type $tp$. Let $n_1, n_2$ be two arbitrary nodes of a process tree, Path $(n_1, n_2)$ refers to the path from $n_1$ to $n_2$ (excluding $n_1$ and $n_2$). Path $(n_1, n_2) = \emptyset$ if $n_2$ is not reachable from $n_1$ or $n_1 = n_2$. Referring back to Fig. 1, Path (Node 0, Node 8) = <Node 2>, Path (Node 1, Node 8) = $\emptyset$.

**Definition 4 (Switch process tree and switch behaviour).** Assume a finite alphabet $A$ of activities. A switch process tree is a normal process tree with switch leaf operators $a \Rightarrow B$ where $a \in A, B \subset A$. Combined with an exclusive choice operator X, the novel leaf node denotes the place we execute activity $a$, and have an option to switch to one of the activities in set $B$ on another branch of an exclusive choice operator. $a \Rightarrow b$ is a switch behavior if there exists $a \Rightarrow B$ such that $b \in B$, we call $a$ the source of the switch behavior, $b$ the destination of the switch behaviour. To ensure the process model is still sound, we define the constraints below:
 1. The activities on different sides of a switch leaf node must be put on different branches of an exclusive choice operator X, i.e. we can only switch execution rights from one exclusive choice branch to another.
 2. If there exists a leaf operator $a \Rightarrow B$ in the process tree, then $\forall b \in B$, First $(a \Rightarrow B, \wedge)$ = First $(b, \wedge)$, and if First $(a \Rightarrow B, \wedge)$ = First $(b, \wedge) \neq \emptyset$, then Path (First $(a \Rightarrow B, \wedge), a \Rightarrow B$) $\cap$ Path (First $(b, \wedge), b) \neq \emptyset$. i.e. we cannot switch out of a parallel branch.

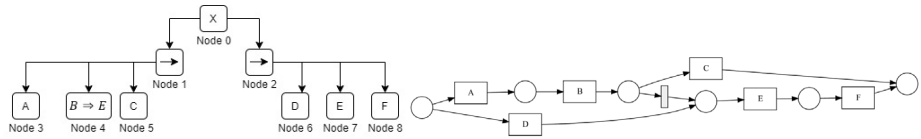

**Fig. 1.** An example switch process tree and its corresponding workflow net

Fig. 1 shows an example switch process tree and its corresponding workflow net. There are three possible traces in the model, which are <A, B, C>, <D, E, F> and <A, B, E, F>.

**Definition 5 (Translating switch process trees into workflow nets).** Translating a switch process tree into a workflow net is straightforward. We first ignore the switch leaf nodes and translate the process tree into a block-structured workflow net. Then we connect the activities on different sides of the switch operators using hidden transitions.

Suppose $Tr$ is a switch process tree, $S$ is the set of all the switch leaf nodes in $Tr$, $Tr^*$ is an equivalent process tree of $Tr$ but all the switch behaviours are removed, i.e., for all the $a \Rightarrow B \in S$ in $Tr$, we convert them into $a$ in $Tr^*$. $N = (P, T_v \cup T_{iv}, F)$ is a block-structured workflow net corresponding to $Tr^*$. For each $a \Rightarrow B \in S$ and $b \in B$, we create a new invisible task $t_{switch}$ into set $T_{switch}$, then:

1. If $|a \cdot| = 1$ in N, $p_{a-out} = a \cdot$, $|\cdot p_{a-out} \backslash T_{switch}| = 1$, then we add a new arc $f$ into $N$, $f = (p_{a-out}, t_{switch})$.
2. If $|\cdot b| = 1$, $p_{b-in} = \cdot b$, $|p_{b-in} \cdot \backslash T_{switch}| = 1$, then we add a new arc $f$ into $N$, $f = (t_{switch}, p_{b-in})$.
3. If $|a \cdot| = 1$ in N, $p_{a-out} = a \cdot$, $|\cdot p_{a-out} \backslash T_{switch}| > 1$, we first delete the arc $f_1 = (a, p_{a-out})$ from N, then we create another new invisible task $t_{bridge}$ and place $p_{bridge}$ into N. We finally add arcs $f_2 = (a, p_{bridge})$, $f_3 = (p_{bridge}, t_{bridge})$, $f_4 = (t_{bridge}, p_{a-out})$ and $f_5 = (p_{bridge}, t_{switch})$.
4. If $|\cdot b| = 1$, $p_{b-in} = \cdot b$, $|p_{b-in} \cdot \backslash T_{switch}| > 1$, we first delete the arc $f_1 = (p_{b-in}, b)$ from N, then we create another new invisible task $t_{bridge}$ and place $p_{bridge}$ into N. We finally add arcs $f_2 = (p_{bridge}, b)$, $f_3 = (t_{bridge}, p_{bridge})$, $f_4 = (p_{b-in}, t_{bridge})$ and $f_5 = (t_{switch}, p_{bridge})$.
5. If $|a \cdot| > 1$, there is a "and split" after $a$. We add a new invisible task $t_{bridge}$ after $a$ as the split point and then go back to step 1, i.e., $|a \cdot| = 1$, $|t_{bridge} \cdot| > 1$, $a \cdot \cap \cdot t_{bridge} \neq \emptyset$.
6. If $|\cdot b| > 1$, there is a "and join" before $b$. We add a new invisible task $t_{bridge}$ before $b$ as the joining point and then go back to step 1, i.e., $|\cdot b| = 1$, $|\cdot t_{bridge}| > 1$, $\cdot b \cap t_{bridge} \cdot \neq \emptyset$.

To illustrate the translation process, we use three examples translated from the above different scenarios in Fig. 2 - Fig. 4:

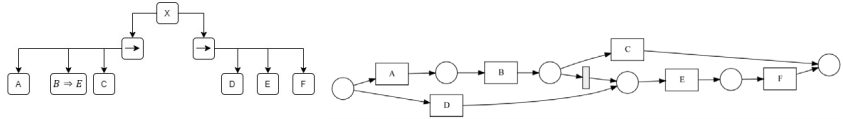

**Fig. 2.** An example translation from a switch process tree to a workflow net (Definition 5, case 1, 2).

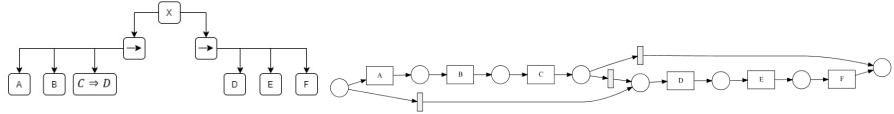

**Fig. 3.** An example translation from a switch process tree to a workflow net (Definition 5, case 3, 4).

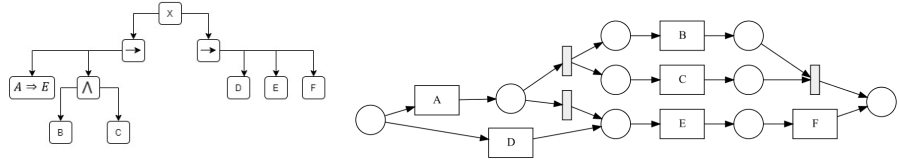

**Fig. 4.** An example translation from a switch process tree to a workflow net (Definition 5, case 5).

**Theorem 1.** If we translate a switch process tree into a workflow net, the resulting workflow net is always sound if the constraints in Definition 4 are all satisfied.

**Proof.** Assume we ignore all the switch behaviours in the process tree during the translation, according to [3], we can get an equivalent sound block-structured workflow net. According to Definition 5, each switch invisible transition is always connected to one single input place and one single output place. The translation process does not increase the number of input/output places of any transitions. As a result, it is free for a token in the model to choose whether firing a switch invisible transition or not. Thus, the resulting process model will not contain dead tasks and is always safe. In addition, since the original block-structured workflow net is sound, if we move a token from one exclusive choice branch to another one, the process can still be completed properly as long as we don't move the token out of a parallel branch. Thus, if the constraints in Definition 4 are all satisfied, the resulting workflow net is always sound.

## 5 Discovering Switch Process Trees

In [6], researchers define the prime invisible tasks into SKIP, REDO, SWITCH, INITIALIZE and FINALIZE where SWITCH refers to switching execution rights between alternative branches. Thus, the SWITCH invisible tasks can be used to represent the switch behaviours we define in Section 4. Researchers in [6] prove that given L is a complete event log of a sound DIWF-net N = (P, $T_v \cup T_{iv}$, F), if $a, b \in T_v$ are two visible tasks, then there is a prime invisible task $t \in T_{iv}$ between $a$ and b, i.e., $a \cdot \cap \cdot t \neq \emptyset$ and $t \cdot \cap \cdot b \neq \emptyset$ iff $a \rightsquigarrow_L b$. Although the scope of the proof is limited, the evaluation of [6] shows that the power of the theory is not limited to complete logs of DIWF-nets. More importantly, [6] provides us with a heuristic to predetermine possible invisible tasks between activities from event logs directly. Suppose we know two activities are on two different exclusive choice branches and there is an invisible task between them, then we know there is a switch behaviour between the two activities.

To discover switch process trees using the inductive miner, we extend the normal exclusive choice cut of the inductive miner framework to a switch exclusive choice cut. In this section, we show the switch exclusive choice cut step by step. To illustrate the process, we use a complete log of the example model presented in Fig. 2 $L_1 = <A, B, C>, <D, E, F>, <A, B, E, F>$ as a running example. To make sure we detect all the switch behaviours, we put the switch exclusive choice cut before the other three cuts in each iteration. The extended IM framework is shown in Algorithm 1.

| | **Algorithm 1. The extended IM framework augumented with switch behaviours** |
|---|---|
| | **Input**: An event log $L$ |
| | **Output**: A Switch Process Tree $Tr$ |
| | $Discover\ (L)$ |
| 1 | **If** $BaseCase(L)\ !=\ \phi$ **Then Return** $BaseCase(L)$ |
| 2 | **Else** |
| 3 | $(cut,(\Sigma_1,\dots,\Sigma_k),switches) = SwitchExclusiveChoiceCut(G(L),L)$ |
| 4 | **If** $k \le 1$ **Then** $(cut,(\Sigma_1,\dots,\Sigma_k)) = SequenceCut(G(L))$ //cut in the original IM |
| 5 | **If** $k \le 1$ **Then** $(cut,(\Sigma_1,\dots,\Sigma_k)) = ConcurrentCut(G(L))$ //cut in the original IM |
| 6 | **If** $k \le 1$ **Then** $(cut,(\Sigma_1,\dots,\Sigma_k)) = LoopCut(G(L))$ //cut in the original IM |
| 7 | **If** $cut = \phi$ **Then Return** $Fallthrough(L)$ //function in the original IM |
| 8 | $NumOfActivitiesBefore = CountActivities(L)$ |
| 9 | $(L_1,\dots,L_k) = SplitLog(L,(cut,(\Sigma_1,\dots,\Sigma_k)))$ //function in the original IM |
| 10 | **If** $cut = SwitchExclusiveChoiceCut$ **Then** |
| 11 | $NumOfActivitiesAfter = CountActivities(L_1,\dots,L_k)$ |
| 12 | **If** $NumOfActivitiesBefore\ !=\ NumOfActivitiesAfter$ **Then** |
| 13 | $(cut,(\Sigma_1,\dots,\Sigma_k)) = XORCut(G(L),L)$ //Exclusive choice cut of the original IM |
| 14 | **If** $k \le 1$ **Then** $(cut,(\Sigma_1,\dots,\Sigma_k)) = SequenceCut(G(L))$ //cut in the original IM |
| 15 | **If** $k \le 1$ **Then** $(cut,(\Sigma_1,\dots,\Sigma_k)) = ConcurrentCut(G(L))$ //cut in the original IM |
| 16 | **If** $k \le 1$ **Then** $(cut,(\Sigma_1,\dots,\Sigma_k)) = LoopCut(G(L))$ //cut in the original IM |
| 17 | **If** $cut = \phi$ **Then Return** $Fallthrough(L)$ //function in the original IM |
| 18 | $(L_1,\dots,L_k) = SplitLog(L,(cut,(\Sigma_1,\dots,\Sigma_k)))$ //function in the original IM |
| 19 | **Return** $cut(Discover(L_1),\dots,Discover(L_k))$ |

## 5.1 The Switch Exclusive Choice Cut (Line 3)

**Step 1: Adding Artificial Start and End Activities**

According to Definition 5, if the source activity of a switch behaviour is at the end of an exclusive choice branch or if the destination activity of a switch behaviour is at the beginning of an exclusive choice branch, we need to add an extra invisible task before the destination activity or after the source activity to represent the process precisely. However, the process model is no longer a DIWF-net after adding the extra invisible task according to [6], so we may fail to detect an invisible task between the two activities using the mendacious dependency.

To solve the problem, before a switch cut, we first identify all the start and end activities in the event log and add a unique start and end activity to each of them. For example, after adding artificial activities into $L_1$, we get $L_1^* = <Start_A, A, B, C, End_C>, <Start_D, D, E, F, End_F>, <Start_A, A, B, E, F, End_F>$.

**Step 2: Calculating All the Mendacious Dependencies Between Activities**

We then go through the event log and identify all the mendacious dependencies. Besides, we ignore all the mendacious dependencies containing artificial start and end activities. After the mendacious dependencies are identified, we delete all the artificial start and end activities.

In our example, we get one mendacious dependency, which is $B \rightsquigarrow_L E$.

**Step 3: Finding Switch Exclusive Choice Cut and Switch Leaf Operators**
Firstly, if there is a mendacious dependency between two activities in the directly-follows graph, we replace the edge between them with an invisible edge.

**Definition 6 (Invisible edge).** Given $G(L)$ is a directly-follows graph of event logs L, $A$ is the set of activities in L. $a, b \in A$, there is an invisible edge from $a$ to $b$ in $G(L)$ iff $a \leadsto_L b$.

**Definition 7 (Switch exclusive choice cut).** Suppose $E$ is the set of all edges in $G(L)$, $E^*$ is the set of all invisible edges. A switch exclusive choice cut is a cut $\Sigma_1, \Sigma_2, \ldots, \Sigma_n$ of a directly-follows graph $G(L)$ such that:
1. There are only invisible edges between $\Sigma_{1 \leq i \leq n}$ and $\Sigma_{1 \leq j \leq n}$.
$$\forall i \neq j \wedge a_i \in \Sigma_i \wedge a_j \in \Sigma_j: (a_i, a_j) \notin E \backslash E^*$$

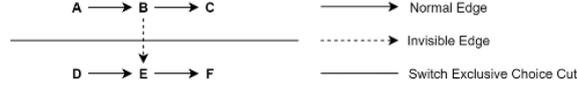

**Fig. 5.** Switch exclusive choice cut for $L_1$

If an invisible edge is cut during the switch exclusive choice cut, i.e., $\exists a_i \in \Sigma_i, a_j \in \Sigma_j: (a_i, a_j) \in E^*$, then a new switch behaviour $a_i \Rightarrow a_j$ is discovered. By merging all the switch behaviours with the same source together in the end, we can get switch leaf operators.

**Step 4: Removing Traces with Switch Behaviours**
We use the same exclusive choice cut split function as the inductive miner infrequent [14] to split the event logs after an exclusive choice switch cut. A problem here is splitting the event log could cause extra "skip" behaviours. In our running example, since we partitioned the activities into two groups which are $\{A, B, C\}$ and $\{D, E, F\}$, the trace <A, B, E, F> will be projected into either <A, B> or <E, F>. The options will either produce an extra end activity B or an extra start activity E in the local sub-process. To resolve the issue, we consider deleting the traces with switch behaviours before splitting the log. For example, we delete <A, B, E, F> from $L_1$ before splitting the log. However, deleting traces increases the requirement of log completeness, which may cause the loss of activities or behaviours when dealing with real-life data. We decide to make the option adjustable.

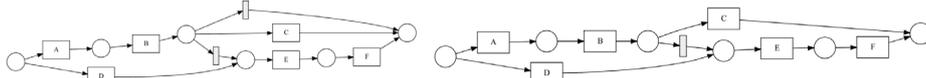

**Fig. 6.** The resulting model of $L_1$ with the deleting traces option off (left) and on (right), there is one more extra trace <A, B> on the left graph.

### 5.2 Verifying the Exclusive Choice Switch Cut (Line 10 - 18)

Performing the switch exclusive choice cut and splitting the event log may cause unnecessary loss of activities. Every time after we perform the switch exclusive choice cut and split the event log, we check if the total number of activities changes. If there is a change in the number of activities (line 12), we abort the whole cut, redo the log split and disable the exclusive choice cut in the next iteration (line 13). We enable the exclusive choice cut again after the current log has been split into sub logs.

### 5.3 Removing Incorrect Switch Behaviours

Although we can identify switch behaviours during the exclusive choice cut, we are unable to determine if the constraints in Definition 4 are met before the whole process tree has been constructed. To ensure a sound model is returned, we iterate through the whole process tree at the end and delete any switch behaviours which violate the constraints defined in Definition 4.

## 6 Evaluation

We implement our approach on the inductive miner directly in the ProM framework [20]. Our code and evaluation results are available at https://github.com/bearlu1996/switch. We applied both artificial and publicly-available event data to evaluate our algorithm. Fitness and precision are used to evaluate the accuracy of our process models. Besides, we use the formula in [4] and [7] to calculate F-score, i.e., $F - Score = 2 * \frac{fitness * precision}{fitness + precision}$. CFC (the number of branching caused by split gateways) [8] and size (the number of nodes) are also used to evaluate the complexity of our process models. For replicable purposes, we use "Replay a log on Petri net for conformance analysis" in ProM to calculate fitness, "Check Precision based on Align-ETConformance" to calculate the precision. We use "Calculate BPMN Metrics" to calculate model complexity. In addition, the tools in the "BPMN Miner" are used to covert between Petri-nets and BPMNs. We use default settings for all the parameters.

### 6.1 Evaluation Using Artificial Data

We first use several artificial logs with switch behaviours to demonstrate the performance of our approach. When applying the original inductive miner on these logs, it fails to discover precise models. Instead, "flower" models with low precision are returned. We show that after using our extension, we can get precise models.

**Table 1.** Evaluation using artificial data

| Log: <A, B, C>, <D, E, F>, <A, F> |
|---|

| IM | IM augmented with switch behaviours |
|---|---|
| 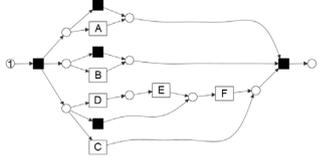 | 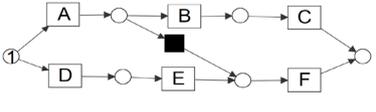 |
| Fitness: 1.0, Precision: 0.38 | Fitness: 1.0, Precision: 1.0 |

**Log: <A, B, C>, <D, E, F>, <A, E, F>, <D, E, C>, <A, E, C>**

| IM | IM augmented with switch behaviours |
|---|---|
| 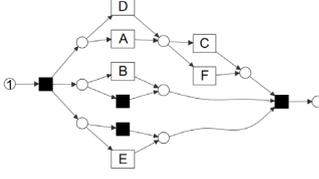 | 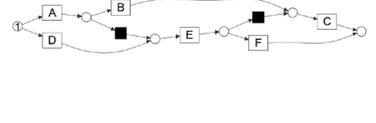 |
| Fitness: 1.0, Precision: 0.42 | Fitness: 1.0, Precision: 1.0 |

**Log: <A, B, C>, <D, E, F>, <A, B, D, E, F>**

| IM | IM augmented with switch behaviours |
|---|---|
| 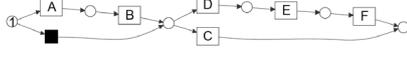 | 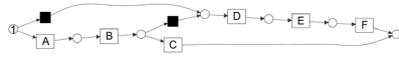 |
| Fitness: 1.0, Precision: 0.94 | Fitness: 1.0, Precision: 1.0 |

**Log: <A, B, C>, <D, E, F>, <D, E, C>, <A, F>, <A, C>**

| IM | IM augmented with switch behaviours |
|---|---|
| 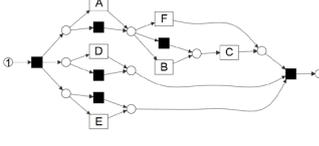 | 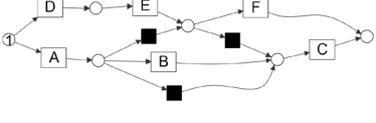 |
| Fitness: 1.0, Precision: 0.42 | Fitness: 1.0, Precision: 0.97 |

### 6.2 Evaluation Using Publically-Available Data

We use a publically-available dataset called "BPIC13-incident" from the "4TU Center for Research Data" to evaluate our algorithm. We use "Event name + lifecycle" as the activity classifier, the dataset contains 7554 traces, 2278 distinct traces, 65533 events and 13 distinct events. The average length of traces is 9 while the shortest length is 1 and the longest length is 123. We combine our approach with the inductive miner infrequent (IMf) [14] and switch off the "delete trace" option, we also compare our results with the split miner (SM) [7]. In addition, we use default settings for all the parameters.

**Table 2.** Evaluation results with the publicly-available dataset (IMs refers to our approach)

|     | Accuracy | | | Complexity | |
| --- | --- | --- | --- | --- | --- |
|     | **Fitness** | **Precision** | **F-Score** | **Size** | **CFC** |
| IMf | 0.95 | 0.59 | 0.73 | 35 | **33** |
| SM | **0.98** | 0.71 | 0.82 | 39 | 48 |
| IMs | 0.97 | **0.80** | **0.88** | **33** | 46 |

Evaluation results are presented in Table 2. All three methods can produce a model with high fitness. However, the IMf returns a model with low precision. Our approach rises the precision of IMf by 36%. In addition, our approach returns a model with both higher precision and F-score than the split miner. For the model complexity, our approach also achieves both smaller size and CFC than the split miner.

## 7      Discussion and Conclusion

In this paper, we present an extension to both the inductive miner and the process tree model. We allow the inductive miner to discover sound process models but not being limited to block-structured workflow nets. The evaluation results show that our approach can reduce the chance for the inductive miner to return flower models. Besides, in our evaluation, our approach can also discover models that are comparable in terms of both model accuracy and complexity to these produced by the split miner.

One limitation is that when performing the switch exclusive choice cut, we do not know if the switch behaviour is valid or not, thus we need to check the validity of the switch behaviours to make sure the model is still sound in the end. It has to be noted that the fitness of resulting models might be reduced if too many switch behaviours are removed. We aim to develop better algorithms to repair the models in the future. Besides, as shown in the artificial data evaluation, when the same place is both the input and output of two switch invisible transitions, there might be redundant hidden transitions in the model, future work is required to remove these redundant hidden transitions.

Finally, we also aim to conduct more experiments to evaluate the performance of our approach in the future, including the impacts of different orders of cuts.